\documentclass{interact}

\usepackage{natbib}
\bibpunct[, ]{(}{)}{;}{a}{}{,}% Citation support using natbib.sty
\usepackage{amsmath,amssymb}
\usepackage{graphicx}
\usepackage{verbatim}
\usepackage{hyperref}

\graphicspath{{./}{fig/}}

\newcommand{\rpkg}[1]{{\bf #1}}
\newcommand{\code}[1]{{\tt #1}}
\newenvironment{api}%
  {\verbatim\leftskip=1cm} {\leftskip=0pt\endverbatim}
\newenvironment{console}%
  {\par\small\verbatim}{\endverbatim\normalsize\par\noindent}

\begin{document}

\title{Modality Analysis via Spacing with the \rpkg{Dimodal} Software Libraries}
\author{
  \name{Greg~Kreider}
  \thanks{CONTACT Greg Kreider.  Email: gkreider@primachvis.com}
  \affil{\small Primordial Machine Vision Systems, Lyndeborough, NH, USA} }

\maketitle

\begin{abstract}
Spacing, the difference between consecutive order statistics, has two
features that reflect the modality of the data.  Consistent, stable
values occur around modes while local increases mark the transitions
between them.  These features not only signal multi-modality, they
also locate modes and anti-modes.  \rpkg{Dimodal} is an R
package for detecting and evaluating these situations.  It includes
parametric feature models and bootstrap tests for spacing smoothed
by low-pass filtering, non-parametric runs and permutation tests
for the interval spacing, and a fusion of changepoints in the raw
spacing.  We introduce the analysis, describe the package, its
implementation and performance, and apply it to identifying
Kirkwood gaps in the asteroid belt.  We also present ports of the
software, with \rpkg{DimodalCPy} a command-line program written in
C with a Python interface.
\end{abstract}


\section{Introduction}
\label{sec:intro}

Imagine a histogram of data and trying to determine if the data it
represents comes from one or more variates.  Bumps of higher counts
indicate modes while transitions between them create dips or gaps at
the anti-modes.  A large count in one of the bins means a smaller
separation between the data points within, assuming they are scattered
over the bin's fixed width. A smaller count at a dip means a larger
distance.  These separations are called the spacing, and the two
behaviors, regions of consistently small spacing separated by local
increases, allow us to determine not only if data is multi-modal, but
where the modes lie.

Our exploration of the use of spacing to analyze multi-modality begins
with a qualitative description of the theory, defines the features, and
sets out seven tests to evaluate their significance.  We then show how
to use \rpkg{Dimodal}, the R reference implementation of this
work, and describe its coding and performance.  This is the basis of
\rpkg{DimodalC}, a port into C, and \rpkg{DimodalPy}, a Python module
build atop the C implementation.  These versions simplify the detectors
and tests where possible and remove one class of test which uses
external R libraries.  Finally, we present a practical example, looking
at the orbital axis of asteroids to identify Kirkwood gaps at the
anti-modes and potential families at the modes.

\section{Modality analysis}
\label{sec:modanal}

\subsection{Spacing}
\label{ssec:moddi}

Using Pyke's notation \citep{pyke65}, spacing is defined as
$ D_{i} = T_{i} - T_{i-1} $, where $ T_{j} $ are order statistics and
the index $ i $ runs from 2 to $ n $, the length of the data.  Its
density is
\begin{equation} \label{eq:fdi}
f_{D_{i}}(y) = \frac{n!}{(i-2)! (n-i)!}
  \int_{-\infty}^{\infty}
    \left\{ F(x) \right\}^{i-2} \left\{ 1 - F(x+y) \right\}^{n-i}
    f(x) f(x+y) \; dx
\end{equation}
where $ f $ is a density function and $ F $ the corresponding
distribution function or c.d.f.  Only a few distributions --- uniform,
exponential, logistic, and Gumbel --- can be integrated to give a
closed-form expression for the density.  The expected spacing is the
first moment over all positive $ y $,
\begin{equation} \label{eq:edi}
E\Bigl\{ D_{i} \Bigr\} = \int_{0}^{\infty} y~f_{D_{i}}(y)~dy
\end{equation}
For logistic variates with scale parameter $ \sigma $
\citep[(17)]{kreider23},
\begin{equation} \label{eq:edi.logis}
E\Bigl\{ D_{i,logis} \Bigr\} = \frac{\sigma n}{(i-1) (n-i+1)}
\end{equation}

\begin{figure}
\centering
\begin{minipage}[t]{\textwidth}
\includegraphics[width=\textwidth]{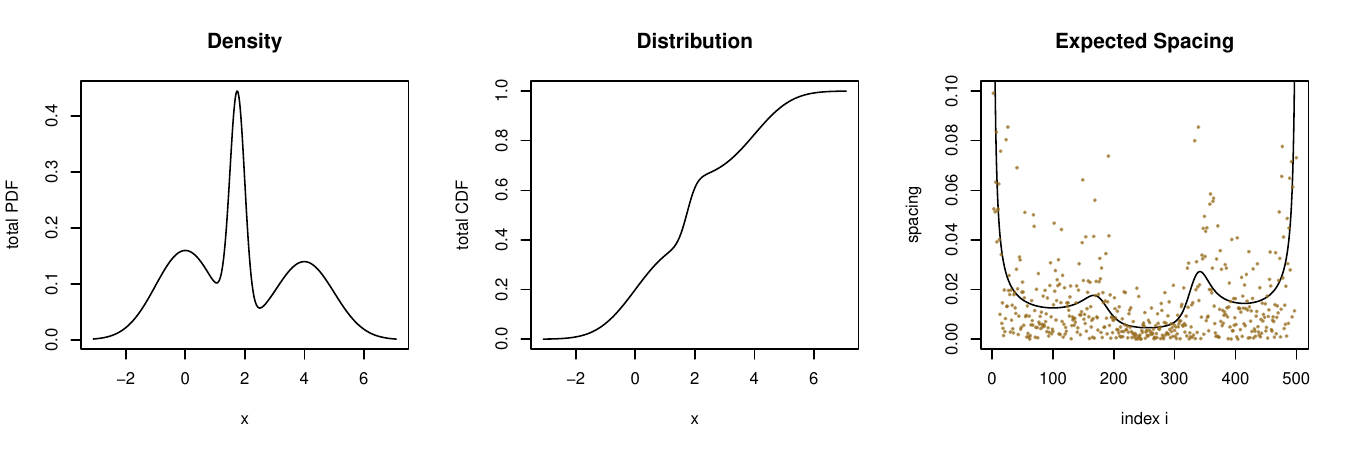}
\caption{\label{fig:edi} Example of spacing in a tri-modal setup. }
\end{minipage}
\end{figure}

The shape of \eqref{eq:edi.logis} is typical of two-sided distributions.
There is a broad span where the expected spacing changes little, with a
third of the points staying within 10\% of the minimum at $ i = n/2 $.
The spacing rises sharply in the tails, becoming $ n / 4 $ times larger
than the minimum. Overall the curve resembles a `U' with a wide floor.
It is these two behaviors that we want to find in multi-modal
situations.  Within a mode the spacing will be stable, with a value
that depends on the distribution's parameters, here $ \sigma $, and
over a length that depends on the size of the draw. Between modes the
spacing will increase by an amount that depends on their separation
and the shape of their tails.  Overall the multi-modal spacing will
still resemble a `U', with the outer tails dominant and much smaller
features appearing in the middle.  Unfortunately, although it is
easy to combine many variates for $ f $ and $ F $ by forming a
weighted sum of the individual density and distribution functions,
\eqref{eq:fdi} and its moments cannot be solved analytically and we
are unable to directly analyze multi-modal spacing.  Trying to model
the features also fails because the interaction is not polynomial
in the variate parameters.  Figure~\ref{fig:edi} plots the density
and distribution functions for a combination of three normal
variates,
\begin{equation} \label{eq:triex}
200 \times N(0,1), \quad 125 \times N(1.75, 0.75), \quad 175 \times N(4,1)
\end{equation}
where the second parameter is the standard deviation.  The expected
spacing, the line in the right graph, is \eqref{eq:edi} integrated
numerically.  Its $ y $ axis is clipped to show the three modes and
two peaks; there are three points at each side that rise outside the
upper limit.  The expected spacing ranges from 0.00459 to 0.337, a
growth of $ 73 \times $.  It is visually flat within the second
variate, and also achieves a broad minimum in the first.  The trailing
tail to the right rounds off the third mode.  Peaks lie at indices 165
and 340, or raw values 0.93 and 2.66, which compares to the anti-modes
in the density at 1.05 and 2.49.  The different draw sizes of the first
and third mode explain most of the shift, with a second-order correction
from the asymmetric placement of the second variate.  The points in the
right graph are the spacing of a sample draw from \eqref{eq:triex}.
Increases exist around the expected peaks and the second variate
produces consistent smaller spacings, but there is a wide spread in
values in the first and third modes.

Dealing with this high variance is the challenge when working with
the spacing.  For exponential variates it equals the square of the
expected value, and simulations imply that this is also true for
logistic variates, although we have not simplified the series
solution for $ V\left\{ D_{i,logis} \right\} $.  A natural solution
is to run the spacing through a low-pass filter and to analyze the
smoothed signal.  \citep{harris78} recommends a Kaiser filter, and
our analysis during development seconds that \citep{kreider24}.
Filtering has the advantage of allowing the analysis of discrete
variates or data taken to only a few decimal places, where ties
are common and the spacing is non-zero only at the points where
the value changes.  The filter smooths these steps, responding to
their density.  Another approach is to use the interval spacing
$ D_{i,w} $ over a larger span than adjacent order statistics.
Extending Pyke's notation,
\begin{equation} \label{eq:diw}
D_{i,w} = T_{i} - T_{i-w}
\end{equation}
where index $ i $ is the upper end of the interval and $ w $ its
width; $ w < i \le n $.  The interval spacing is the sum of the
individual spacings over the width.
\begin{align} \label{eq:diwsum}
\sum_{j=0}^{w-1} D_{i-j}
  & = (T_{i} - T_{i-1}) + (T_{i-1} - T_{i-2}) + \ldots
    + (T_{i-w+2} - T_{i-w+1}) + (T_{i-w+1} - T_{i-w}) \nonumber \\
  & = T_{i} - T_{i-w} = D_{i,w}
\end{align}
Such a sum is the same as applying a rectangular or running mean filter
to the spacing, albeit without normalizing by the width so that the
signal will appear to be amplified.  This is usually considered a poor
filter because it has little sidelobe suppression and high frequency
components will pass through. The interval spacing will be rough.  The
wide mainlobe means the interval can be smaller than other low-pass
kernels.

\subsection{Features}
\label{ssec:modfeat}

\begin{figure}
\centering
\begin{minipage}[t]{0.75\textwidth}
\includegraphics[width=\textwidth]{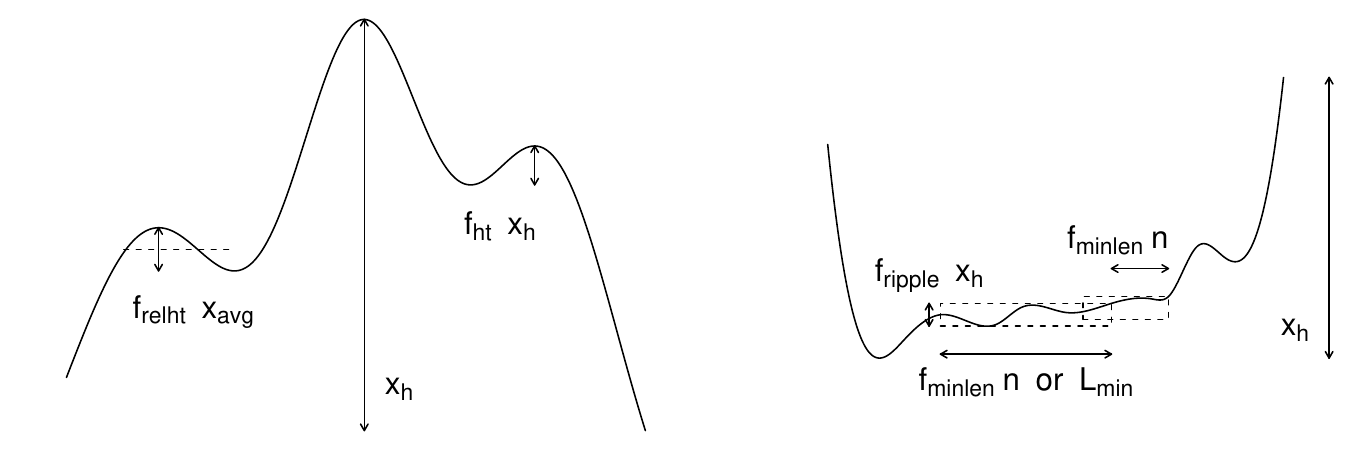}
\caption{\label{fig:featdef} Screening requirements for peaks (left) and
flats (right).}
\end{minipage}
\end{figure}

Within the processed signal, either the spacing after low-pass filtering
or over intervals larger than one, we want to look for two features.
Peaks are local maxima that have been screened to eliminate small
subsidiaries to the side of larger increases.  The detector first
compresses series of nearly equal values to a single point.  This has
the advantage of allowing some noise at the feature, which is useful
when working with the interval spacing, and turns an ideal square wave
into alternating high and low points.  It then identifies the extrema,
those points that are larger or smaller than both neighbors.  It
measures the height of the maxima above the neighboring minima.
Starting with the smallest, the detector merges a maximum into its
neighbor if the intervening minimum is shallow, defined as a height
less than some fraction $ f_{ht} $ of the range of the data or a
fraction $ f_{relht} $ of the average of the two
(Figure~\ref{fig:featdef}).  Flats, the second feature, are sequences
of similar data values.  They are defined by a ripple parameter
$ f_{ripple} $ as a fraction of the data range, and we add a minimum
length requirement, both in absolute terms $ L_{min} $ and as a
fraction of the amount of data $ f_{minlen} $.  Flats are found by
determining the ripple range at each data point and scanning outward
until the data leaves those bounds.  To account for local noise the
detector can skip a limited number of points that fall outside the
ripple, where typically $ n_{outlier} = 1 $. The detector keeps the
longest flat at each point; if a shorter flat projects beyond a
longer by a length meeting the requirements, it too is kept
(Figure~\ref{fig:featdef}, rightmost flat).  Flats may therefore
overlap.

%% !!!
% Generate new version without xsf.  Should be frel * xavg.  Less height.
% Generate flat, fuzzy run (?).

A third feature detector, for fuzzy runs, is used internally in several
places.  This is a traditional run length encoding for discrete data,
including integers, characters, and symbols, but it treats real values
as equal when their relative difference, the same delta-over-average
fraction used to screen peaks, is below a threshold, typically small.
It tracks and skips invalid (NaN) values.  The detector scans the data,
running forward from each point while the criterion is met, and
chooses the first point to represent the entire run. Runs are used for
the initial compression in the peak detector, for characterizing peaks
in the interval spacing, and when approximating distribution functions
with piecewise linear curves.

\subsection{Evaluation}
\label{ssec:modtest}

Having identified peaks and flats, the next step is to evaluate the
features to determine if they represent real changes in the spacing.
We use three broad strategies: modeling the probability of features
arising in the low-pass spacing assuming a null univariate
distribution; checking runs of increasing or decreasing or tied
values in the interval spacing; and bootstrap or excursion tests
based on either spacing.  The tests were developed in this order to
minimize assumptions about the data.  They do not have a strong
correlation with each other and can be used as independent checks
on a feature's significance.  A fourth strategy will look for
changes in the behavior of the spacing without any filtering.

The models for peak height and flat length are built by filtering
the spacing of draws from a single base variant and collecting the
detected features.  The models link the quantiles of these values
to the parameters: the draw size $ n $ varying between 50 and 500;
the filter size $ f_{lp} $ between $ 0.05 \, n $ and $ 0.40 \, n $;
the filter; and the base distribution.  The models do not have a
theoretical basis and were picked for their fit to the collected
data.

The height model is complex but is built on a stable foundation.
There is a conservative choice for a null distribution, an asymmetric
Weibull with scale parameter $ a = 2 $ and shape parameter $ b = 4 $.
It gives larger critical values than other variates but over a narrow
range, with its 0.95 quantile height matching the 0.975 or 0.99
quantile from other distributions.  Uniform and beta variates are not
included, as they generate many more and larger peaks and give
critical values beyond the 0.999 quantile of the others.  The Kaiser
filter produces fewer and smaller peaks, supporting its choice as the
preferred kernel.  Other filters have critical values that are roughly
proportional to the Kaiser over all quantiles of interest, so they can
be handled by scaling.  A preliminary fit with the null variate and
filter showed the height varied with $ \log n $ but that interactions
with $ f_{lp} $ and the quantile $ q $ could not be captured.
Instead, the quantiles best follow a Wald or inverse Gaussian
distribution whose location and scale parameters, and a scaling of
the height, are fit by regression. This would be the correct
distribution if the peak were a Brownian excursion, although that
cannot be true if only because the spacing cannot be normally
distributed since it is always positive.  Critical values follow
from the Wald quantile function with the same parameter fits and an
inversion of the height scaling.  The model fits the simulated
draws within 5\% and degrades for small samples $ n \le 70 $ or
filter sizes $ f_{lp} \le 0.075 $, or with large filters
$ f_{lp} > 0.30 $.

The length model is simpler.  The critical length follows directly
from a regression linear in $ f_{lp} $, quadratic in $ n $, and
quadratic in $ q $ plus a logistic term.  The quantile is found
numerically with a root solver, inverting the model.  The error is
smaller than for the peak model, within 5\% for all parameter
combinations, although its accuracy decreases if $ n \le 60 $
or $ f_{lp} \le 0.075 $.  However, the choice of a null distribution
is not clear.  There is less overlap between the critical values
from different variates and a larger range within one.  The 0.95
quantile from one might apply at the 0.90 or 0.999 quantile in
another.  As a compromise the model adopts a logistic variate as the
base distribution, but offers Weibull and Gumbel options as more
liberal and conservative alternatives.  It also includes an option
for a Gaussian base given the common assumption of normality of
data, although its critical values fall between the logistic
and Weibull.  The Kaiser filter is still preferred, but a simple
scaling of quantiles for other filters does not work and separate
models are needed for all combinations of the base distribution
and filter.

The interval spacing is rough enough that there are runs in its
signed difference.  These are series of increasing and decreasing
values and have been used in trend analysis, for example to identify
if an economy is expanding or contracting.  A peak can form from a
single large step or a run of smaller, and by using the signed
difference we remove the step size and focus only on the trend.  We
can test the number of runs within the feature, or the length of the
longest, or how often permutations of them re-create the peak.  The
general combinatorial analysis of \citep{kaplansky45} shows the count
is distributed normally with simple expressions for the mean and
variance.  The runs count test compares the actual number within the
peak against this distribution.  The analysis assumes the symbols
--- increasing, decreasing, or steady --- are independent, however,
and if we want to account for the correlations introduced by the
overlap of adjacent intervals we can model the signed difference as
a Markov process.  After measuring the transition rate between
symbols over all the data, we split the matrix into two parts, the
diagonal that advances each symbol to itself along a run, and the
rest that start a new run.  Using these sub-matrices to set up a new
transition matrix for the growth of a run in any symbol, we can
determine the probability of its length by stepping the matrix over
the size of the feature and seeing if the chain reaches that length,
then weighting over the states by the initial or steady state
distribution to account for all symbols.  The stepping reduces to a
recursion using the two sub-matrices \citep{kreider25b}.  The last
test permutes the set of runs and regenerates the feature.  The
quantile of the height is the fraction of permutations that yield
at least as prominent a peak, which require that long rises and
falls lie on different sides of the maximum, interrupted only by
short moves in the other direction.  These interruptions amount to
a restriction on the permutation that the run directions must
change.  Runs in the same direction cannot be placed adjacent or
they would form a longer run than what exists in the data.  To
avoid permutations that reconstruct a minimum or valley, the
simulated height is defined as the signal's range above the lower
of the first or last point bounding the feature, a definition
which mirrors the actual measurement. Of course there will usually
be too many permutations to exhaustively check, and sampling will
be needed.

Following the spirit of the permutation test, instead of working with
runs we can look at how often peaks occur when drawing from a pool
formed by the difference of the low-pass or interval spacing, or in
other words, its derivative. This is a bootstrap test where the pool
is sampled with replacement, pulling as many data points as compose
the feature and performing a cumulative sum to re-create it.  We call
it an excursion test, although the feature need not return to its
initial value at the end.  The feature's quantile is the number of
bootstrap samples that give at least as great a height as a peak. The
test can simulate flats too, whose reconstructed height is now
defined as the range of the excursion, matching the ripple
specification, and whose quantile follows from the number of samples
that are smaller.  In a way a flat excursion inverts the model,
taking the length as given and comparing the height.

The tests of peaks in the low-pass spacing do well.  They are a little
more sensitive than existing uni-modality checks.  The height model
accepts smaller differences than the excursion test, although the
inaccuracies in the fit mean it should be evaluated at a tighter
passing level, at which point the performance is the same.  The
excursion test applied to peaks in the interval spacing is less
sensitive, but this counters the larger number of features found in
the rougher signal and the number passing is the same.  The runs
count and longest run tests perform the same, despite the different
assumptions behind them.  They are very insensitive and slow to
respond, and need tight acceptance thresholds to avoid a large false
positive rate.  A passing peak in these tests is a strong sign for
multi-modality but will likely duplicate other test results.  The
run height permutation test is as sensitive as the low-pass tests
but it readily accepts marginal peaks, even with a very tight
acceptance threshold.  Passing peaks do locate changes in modality,
with more uncertainty in their position when found in the interval
spacing.

The tests of flats in either spacing perform the same.  They are
selective, accepting one out of five detected.  Those that pass cover
a mode, although the endpoints are usually not stable.  The middle of
a flat is an accurate estimator of the mode's location, confirming
\citep{venter67}.

\subsection{Changepoints}
\label{ssec:modcpt}

For the last analysis we draw on the rich history of changepoint
detectors, which reaches back to the development of statistical
process control techniques and economic trend analysis.  They are
found in more than 25 maintained R packages, and some implement
more than one detection strategy.  They use a variety of
algorithms, including: two-sample tests to find where the mean or
variance changes; cumulative sums reaching control limits that
handle trends; partitioning based on information or entropy
criteria; consistent regression fits; and identifying where updates
to Bayesian priors are needed.  In our experience no detector is
robust and trustworthy in all situations, and the quality of the
results and stability of the implementation vary widely.  Detectors
can be noisy and inconsistent in their placement of changepoints,
or can miss changes.  For example, several algorithms are unable
to handle discrete variates, firing at each new value.  Other
algorithms require a burn-in period to characterize the local
statistics and cannot respond quickly to narrow features.  Others
delineate regions without changes and not single points.  Some
detectors are expensive and cannot handle more than a few hundred
data points, and others are unstable, in the worst case crashing
the R session.

Changepoints respond to transitions between the modes and anti-modes.
If these are sharp, such as the second mode of \eqref{eq:triex},
they can locate the edge of features, but usually they will instead
bound and separate them.

Few changepoint algorithms provide a significance level to their
results, instead taking an acceptance level or a proxy, such as the
average run length, as a decision parameter.  This restricts the
options to combine their results.  With just a yes/no decision on
each data point and no test statistic, the only classifier fusion
strategy left is majority voting \citep{ruta00}.  The changepoint
analysis runs any detectors it finds on the system and passes those
points that appear in most of the results.  It does some screening
for consistency, ignoring detectors that are unusually quiet or
noisy, and allows for some misalignment of points by considering
those nearby to be the same.  Majority voting handles the
inconsistencies in individual detectors to extract the stable
points they have in common.

By its nature, changepoint analysis complements the feature detectors.
It finds the flanks of peaks or flats, often on just one side.  It
cannot judge the significance of any point, and adding such a test
would only duplicate the work of one of the algorithm, undoing the
fusion.  We recommend a minimal set of packages: the non-parametric
PERT library \rpkg{changepoint.np} \citep{haynes22}, the Iterative
Cumulative Sum of Squares package \rpkg{ICSS} \citep{koestlmeier21},
and the joint segmentation library \rpkg{jointseg} \citep{pierre19}.
They have three complementary approaches, involving the partitioning
of data according to an information criteria penalty function,
process control bounds monitoring, and consistent regression fits.
The analysis must be able to handle any detectors available on the
system, however.

\subsection{Summary}
\label{ssec:modsum}

In summary, detecting modality in the spacing of data can be done by
finding features, peaks from local increases at the anti-modes and
flats from consistent sections around the modes. These features are
easiest to see after smoothing the spacing, finely with a low-pass
filter or coarsely using the interval spacing.  Parametric models
of the low-pass features based on an assumed distribution of the
original data offer one way to evaluate their significance.  The
peak models based on a conservative choice of null distribution
have a somewhat loose fit to simulated draws and require a
conservative acceptance level, while the flat models fit better but
are sensitive to the choice of the base distribution.  Within the
interval spacing the number or maximum length of runs of increasing,
decreasing, or tied spacing within a feature provide a second set
of tests.  A third runs test counts the permutations that re-create
the peak.  This reconstruction can also be done in either filtered
spacing by bootstrap sampling its difference and re-creating the
excursion.  All of these tests complement each other and a positive
result in any should be considered evidence for a mode or
multi-modality.  Finally, changepoints signal the transition between
features, and a majority vote of the wide variety of detection
algorithms allows us to use a large number of complementary
approaches to improve the results.  However, this does not allow
judging the significance of the features, nor does it locate them
unless the transition between modes is sharp.

\citep{kreider25a} describes the detectors and tests in full, providing
pseudo-code and an explanation of all parameters.

These analyses do have some limitations.  They apply only to
one-dimensional data, as higher dimensions have no ordering; this
is not a clustering analysis.  The interval spacing and changepoints
have difficulty handling discrete or quantized data, as they respond
strongly to the step at each change in value.  A low-pass kernel
smooths the steps away, but it must be large enough to span them
without becoming so large that it damps out peaks.  If the number at
each value tails away, for example with Poisson variates, then
there is a balancing act between choosing a filter size that
responds to the large counts at one and small counts at another.  A
similar problem arises with extremely large data sets with tens or
hundreds of thousands of points, where the spacing without
filtering is very smooth and features exist on a large scale.  A
mode must contain enough points to support a flat, and smoothing
by the filter will soften the edges and shrink the stable region.
Perhaps most importantly, the initial and final tails in the
spacing --- the arms of the overall `U' --- will hide changes, making
the detection of peaks at the edges difficult and of flats almost
impossible.  The tails need small filter kernels to resolve the
rapid change, but it is here that the feature models start to break
down.  The interval spacing has better success.

\section[The Dimodal package]{The \rpkg{Dimodal} package}
\label{sec:Dimodal}

\subsection{Interface}
\label{sec:Rapi}

\rpkg{Dimodal} \citep{kreider25c} is an R package that analyzes the
spacing for multi-modality.  It implements the feature detectors
and significance tests of Section~\ref{sec:modanal}.  (The version
posted to CRAN omits the changepoint fusion, because of limitations
in how that platform deals with optional package dependencies.)  In
principle its use is just
\begin{console}
R> m <- Dimodal(x)
\end{console}
$ x $ is a vector with the data.  $ m $ is an object of S3 class
\code{"Dimodal"}, which is a list with supporting \code{print()},
\code{summary()}, and \code{plot()} methods.  $ m $ will contain up to
seven elements, depending on the analysis run. \code{m\$data} is a
matrix cast to class \code{"Didata"} that has the raw, low-pass, and/or
interval spacing.  \code{m\$lp.peaks} and \code{m\$diw.peaks} are data
frames cast to class \code{"Dipeak"} that contain the local extrema in
the low-pass and interval spacing respectively, and for the peaks their
test statistics and probabilities.  \code{m\$lp.flats} and
\code{m\$diw.flats} are the equivalent \code{"Diflat"} data frames for
flats.  \code{m\$cpt} is a list of class \code{"Dicpt"} with the
changepoints from each detector and the voting that combines them.
Finally, \code{m\$opt} is a list of analysis options that control the
processing.  Each of the data classes supports \code{print()} and
\code{summary()} methods and a \code{mark()} method used to annotate
graphs.  The \code{"Dicpt"} class also includes a \code{plot()}
method that shows the changepoints from each algorithm and the voting
for the final set.

\code{Dimodal()} can take a second argument to change the analysis
parameters.  The \code{Diopt()} function is an options database
modeled on the \code{par()} function.  It can be called in five ways
\begin{api}
Diopt()
Diopt(NULL)
Diopt(key1=val1, key2=val2, ...)
Diopt(list(key1=val1, key2=val2, ...))
Diopt("key1", "key2", ...)
\end{api}

The first form returns a list with all parameters.  It provides the
default values for \code{Dimodal()}.  The second resets the options
to the package default.  The third and fourth change the database
and return a list of the original values of the keys.  Do
not use these versions in a call to \code{Dimodal()}, which expects
the complete set.  Like \code{par()},
\code{Diopt(Diopt(key1=val1, ...))} undoes the inner change,
restoring the state.  The last version returns a list with the
requested keys' values.

To temporarily make a change without affecting the database call
\begin{api}
Diopt.local(key1=val1, key2=val2, ...)
\end{api}
This returns a list of all options with the argument keys overridden.
Use this version in a call to \code{Dimodal()}.

The options include feature detector parameters, acceptance levels
for the tests, voting requirements for the changepoint fusion, and
display colors and formatting.  All told there are 50 values in the
database, described on the command's help page.  For example, the
\code{"analysis"} option tells which analyses to run --- "lp" for
the low-pass spacing, "diw" for the interval, and "cpt" for
changepoints.  The default value includes all three, but to run
just the filtered routines once, use
\begin{console}
R> m <- Dimodal(x, Diopt.local(analysis=c("lp", "diw")))
\end{console}
Calling
\begin{console}
R> oldopt <- Diopt(analysis=c("lp", "diw"))
\end{console}
would mean skipping the changepoint analysis by default in the future.

The package contains one other top-level command that generates a
simplified mode tree \citep{minnotte92}.  These plots show how the
position and significance of peaks and flats vary with the filter size
or bandwidth.  They step the kernel or interval size as a fraction of
the data length from 1\% to 40\% (set by the \code{"track.maxwindow"}
option) and store the features found for each.  No attempt is made to
link features as the bandwidth changes, unlike the mode tree, and the
plot of the result just marks the positions, dots for peaks and bars
along flats, coloring them by the best test probability.
\begin{console}
R> trk <- Ditrack(x, "lp")
\end{console}
Use \code{"diw"} as the second argument to study the interval
spacing.  You can provide a local options override as a third
argument.  The feature tracker is meant to guide the choice of
window size for the analysis by looking at the stability of features
or if they persist over a wide range of bandwidths, which we will
see in the example.  Printing the result shows a table with the
number of passing features for each kernel size.

\rpkg{Dimodal} exports the feature tests.  For the peak and flat
models it provides
\begin{api}
Dipeak.test(ht, n, flp, filter, lower.tail)
Dipeak.critval(pval, n, flp, filter)
Diflat.test(len, n, flp, filter, basedist, lower.tail)
Diflat.critval(pval, n, flp, filter, basedist)
\end{api}
with parameters $ n $ for the data size, $ flp $ the kernel size,
either as a fraction of $ n $ or an absolute integer value,
$ filter $ the name of the low-pass filter to use, and for flats
$ basedist $ the parametric base model. $ lower.tail $ should be
FALSE although the default value for the argument is TRUE to match
R's distribution functions.

The runs tests
\begin{api}
Dinrun.test(x, stID, endID, feps, lower.tail)
Dirunlen.test(x, stID, endID, feps)
\end{api}
convert the data $ x $ into fuzzy runs between each start and
endpoint index (inclusive) while treating points within relative
height $ feps $ as the same, and then perform the test.  The
symbol populations for the Kaplansky\-Riordan test come from the
feature bounded by the starting and ending indices, and the
transition matrix for the Markov chain model comes from all the
data.  $ lower.tail $ should be TRUE to give the probability that
there are not more runs than observed.  It is not needed for the
run length test.

The excursion tests determine the probability of a feature of
height $ ht $ and size $ ndraw $ while drawing from a set of values
$ xbase $, using
\begin{api}
Diexcurht.test(ht, ndraw, xbase, nexcur, is.peak, lower.tail, seed)
Dipermht.test(ht, xbase, nperm, lower.tail, seed)
\end{api}
The draws are repeated $ nperm $ or $ nexcur $ times, with
$ lower.tail $ FALSE giving the probability of a peak or TRUE of a
flat.  The $ is.peak $ argument switches between the two height
definitions; because the permutation test is only for peaks its call
does not need it. $ seed $ if provided and not zero sets up the
random number generator so the results are repeatable.

These tests are run within \code{Dimodal()}.  Each returns a
\code{"Ditest"} structure which stores the test statistic, $p$ value,
and descriptions of the test and arguments.  This class has
\code{print()} and \code{summary()} methods.

The package provides several utility functions for working with the
results. \code{select.peaks()} returns the subset of the
\code{"Dipeak"} extrema with actual tested peaks, dropping minima.
\code{Dimodal} calls \code{shiftID.place()} to move
feature positions from the filtered signal to the data grid and to
add the corresponding raw data values using the mid-quantile
approximation.  \code{center.diw()} accounts for the different
indexing of the interval spacing by further shifting positions
to the middle of the interval so they correspond to the low-pass
spacing and data.  \code{match.features()} identifies the features
that appear in both the low-pass and interval spacing based on
positional matching, either nearby peaks or overlapping flats.
\code{runs.as.rle()} converts the runs detector output into an
\code{"rle"} encoding.  \code{midquantile()} replaces the stepwise
distribution or quantile function for discrete or heavily quantized
data with a piecewise linear approximation (for example,
\citep{ma11}) and is used to smoothly convert steps in the spacing
to raw values.  By default it will not do the interpolation for
continuous data.  \code{Dimodal()} uses it to convert data
indices back to raw values, and a simple version is used for the
quantiles in the run height and excursion tests.

Finally, although not exported, the package does document the five
feature detectors for use elsewhere.  They will work with any signal,
not just spacing.  They must be accessed within the namespace with
\begin{api}
Dimodal:::find.runs(x, feps)
Dimodal:::find.peaks(x, fht, frelht, fhtie, fhsupp)
Dimodal:::find.flats(x, fripple, minlen, fminlen, noutlier)
Dimodal:::find.level.sections(x, alpha, correct)
Dimodal:::find.cpt(x, cptlibs, fncpt.max, timeout, qvote, sep,
                   fsep, libsep)
\end{api}
$ x $ is the data to analyze and the other arguments are
defined in Section~\ref{sec:modanal}.  The peak, flat, and changepoint
detectors return objects of the appropriate class, which
\code{Dimodal()} extends by adding test results and shifting positions
using the \code{shiftID.place()} utility function.  The level section
detector is a changepoint algorithm that uses a combination of
interval spacing to bound flats.  It and the run detector return lists
with the feature information.

\subsection{Implementation}
\label{sec:Rimpl}

The feature detectors have been written in C for speed.  A goal when
creating the package was that it should be able to handle data sets
with hundreds of thousands if not a million points, sizes which were
encountered during development.  This led to changes in the algorithms
used for the detectors and to moving the excursion and permutation
tests into C.  The R side interface is a thin wrapper that does some
data preparation and formats the results into the appropriate data
class.

The run detector is a straightforward scan of the data, combining points
if they meet the matching criterion.  Integer, factor, logical, and
string values must match exactly.  Real values with a relative difference
less than $ f_{eps} $ are considered the same, as are infinities of the
same sign.  Values within the machine's double precision accuracy always
match, even if $ f_{eps} $ is set to 0, because the internal conversion
of data taken to a few digits is never exact and the subtraction for
the spacing can leave a small residual.

Peak detection uses a minheap to store the local maxima, those points
that are greater than both neighbors, ordered by the height to the
higher of the neighboring minimum.  They are processed one by one,
pulling them off the top of the heap and checking against the height
requirements.  Merging into a dominant neighbor involves deleting the
maximum and minimum pair and re-calculating the height of the
neighboring peak, re-positioning it within the heap.  The code is
complicated by treating the first and last data points as extrema,
which creates corner cases for the merging step.

A simple approach to detecting flats is to check each data point by
determining the ripple bounds and scanning outward to each side until
the signal moves beyond them, counting outliers as it goes along.
This fails to scale, however, because the spacing becomes more
uniform as the data set grows and flats cover a substantial fraction
of it; 30--50\% has been seen.  The scan is $ O(nL) $ for $ n $ data
points and flats of length $ L $, and goes to $ O(n^2) $ as the
fraction increases.  To avoid this scaling an alternate detector
uses a pair of segtrees, one minimum and one maximum, to identify
the range of data within each node of the tree.  The end of a flat
is found by moving upward within the tree as long as the range is
within the ripple bounds, and then descending away from the source
to find the first failing point. This repeats as often as the number
of outliers allows.  Building and walking the segtrees adds
considerable overhead to the simple scan and begins to out-perform
it only when there are 5000 to 10000 data points.  By 75 thousand
points there is an order of magnitude improvement, at 250 thousand
a $ 35 \times $ speed-up, and at 500 thousand $ 74 \times $.

After determining the flat around each data point using either
alternative, the detector then picks the longest flat at each point.
This is done by sorting the flats in decreasing length and
intersecting each with the set of uncovered intervals, initially
the entire data.  A flat can either split a larger uncovered
interval into two by removing a center section, or it can overlap
part or all of one, shrinking it.  If the removed or shrunken section
is longer than the minimum threshold, the detector reports the flat
with its original endpoints.  This allows overlaps.

The longest runs test is a straightforward implementation of the
recursion established when evolving the Markov model of the run
progression through the feature.  It is consistently a couple orders
of magnitude faster than doing the matrix operations in R,
mostly by hard-coding them for the $ 2 \times 2 $ or $ 3 \times 3 $
symbol transitions we need.  The implementation does extend to any
number of symbols, however.

The permutation algorithm first places the directional markers, the
$ -1, 0, \text{and} +1 $ of the signed difference, and then freely
shuffles the run lengths within each symbol.  Alternating two symbols
is simple: start with the larger class or pick one at random if they
have the same population.  For three or more symbols the permutation
picks one at random according to its frequency while keeping the
adjacency prohibition, until one achieves the majority.  At this
time the permutation becomes an alternation of the largest class
with a shuffle of all others.  Adjacency within this shuffle is
unimportant because the symbols will be separated by the majority.
Permuting with more than two classes is noticeably slower as the
choice is made element by element until the final inter-weaving.

Excursion sampling is done with the \rpkg{PCG} random number generator
\citep{oneill14}, unless the package has been compiled with
\code{"-DNO\_PCGRNG"}, in which case R's sampling function
is used.  This is done for speed for large data sets when the total
draw size, the number of excursions times the feature size, comes
within an order of magnitude of the 32 bit space and larger width
integers must be generated.  The PCG generator is $ 5 \times $ faster
creating 64 bit random numbers than the built-in sampler, and is as
efficient as R's 32 bit \code{unif\_rand()}.

The mid-quantile approximation for discrete values offers four
different strategies for replacing the steps of the empirical c.d.f.,
or its inverse, with piecewise linear segments.  Instead of a constant
quantile for a range of tied values, which the R function
\code{ecdf()} provides, it generates a ramp.  The mid-distribution
function of \citep{ma11} places the endpoints of each segment at each
data value and the quantile halfway along the step; in the
implementation this is the type 2 approximation.  Type 1 is the
equivalent for the quantile function, placing the endpoint at the
quantile and shifting to half the data grid.  Type 3 combines both,
using a half-grid for both quantiles and data values.  Type 4 draws
the line segments between the mid-point of each run of tied values,
or through the data points when there is no run and they are
distinct.  Types 1 and 2 can create an envelope around the
distribution, although in practice the Type 2 segments lie inside
the stepped version and are close to the Type 3 values.  Types 2
and 3 will not exactly follow the data when there are no runs.
Type 4 does, and has the smallest fitting error.  The
approximation by default switches from type 4 to type 3 if 90\% of
the data contains repeated values.

Determining the probability of the permutation and excursion tests
employs a simple mid-quantile approximation, counting half of the
draws matching the feature's height, without regard to the step
between levels.  The correction is mostly needed for the run height
test because the height is an integer, its range limited, and many
permutations will generate the same value.  The excursion test will
generate few tied heights, or none if the data is continuous.  The
correction improves the reported quantile but cannot remove the
coarse steps between them that come from the tied values.  The
resolution near the test's passing level may be poor.

The challenge with the changepoint analysis is not the voting scheme
but handling all the external packages.  \rpkg{Dimodal} checks whether
a library is installed with \code{requireNamespace()} and accesses
individual tests with the \code{<pkg>::<function>} notation, which
loads a library without putting it on the search path.  A package
may be used in this way without adding it as a dependency of
\rpkg{Dimodal}.  If problems resolving objects within a library's
namespace arise with this approach, explicitly load the package
beforehand with \code{library("<pkg>")}.  To prevent run-away
algorithms --- some are quite slow even with only a few hundred data
points --- the detector wraps each call in \code{setTimeLimit()} and
traps any interrupts it creates, as well as captures any error or
warning messages.  \code{find.cpt()} has a blacklist option to
ignore some installed libraries and by default excludes two that are
unusually slow.  Otherwise all methods and variations within a
library will be used, as documented on the \code{find.cpt()} help
page.  The detector will do some post-processing of the results to
convert intervals into endpoints and to put everything in a common
format.  All libraries run with their default or recommended
parameters, and there is no ability to change this.  Their results
may not be repeatable, even if the random number generator seed is
set beforehand.

The analysis proceeds in four steps.  First, it runs all variants
within a library and combines the individual results, considering
those that are very close to be the same.  If the library identifies
more than some fraction of the data as changepoints it is ignored as
being noisy.  Second, as a consistency check it drops libraries that
have unusually few or many changepoints compared to their peers.
Third, it merges the library points into a master list, using a
looser separation criterion than the intra-library merge.  Finally,
voting occurs, counting the number of libraries matching each master
point.  The algorithm reports those points that appear in the
majority.

\subsection{Performance}
\label{sec:Rperf}

To evaluate the success of this implementation against the goal, we
run the modality analysis on large subsets of the original dataset
behind the Kirkwood gaps example in the next section.  The raw data
contains more than 1.4 million asteroids, even after limiting them
to the main belt.  We take samples of 10 thousand to 750 thousand
points, run the analysis, and gather timing for each top-level step.
The low-pass filter size is set to $ f_{lp} = 0.05 $ and the
interval width to 1\% of the data.  We choose feature parameters
for each sample size to give the same number of peaks and flats.
Repeating this 25 times and averaging, Table~\ref{tbl:Rperf} gives
the median time in milliseconds when drawing 100 thousand samples
and the scaling from this base for other draw sizes.

The table shows that one analysis step, the low-pass filter, has
complexity $ O( n^{2} ) $.  This is unavoidable.  The convolution
is done as a matrix operation rather than in Fourier space and
the kernel size is proportional to $ n $, whence the quadratic
scaling.  Several steps run in linear time, including generating
the spacings, running the detectors, and testing.  Flat detection
is a little worse than linear, implying the feature sizes scale
with the data.  The permutation and excursion tests dominate the
run time.  Even with 750 thousand data points the quadratic
growth of the low-pass filter does not overtake them, running
for 55~s compared to 70~s for the interval excursion tests and
163~s for the run height.  In comparison the detectors and other
tests need 6~s.  The implementation is capable of handling data
sets with a hundred thousand points but will be slow for much
larger.

% measured on nuveena
%  median in ms    10k     50k    100k    250k    500k    750k
%  peak detect
%   Di               0       3       4      15      24      28
%   Diw              0       2       4      17      10      23
%   LP               9     215     870    5345   21269   47776
%   midquantile      0       2       6      14      36      38
%  LP analysis
%   peak detect      2       1       2       4      10      12
%   ht model         2       0       0       0       0       0
%   peak excur      88     424     802    2148    4817    7239
%   flat detect     47     117     247     670    1454    2554
%   len model        0       0       0       0       0       0
%   flat excur     319    1640    3209    8353   19115   31592
%  Diw analysis
%   peak detect      1       6       8       9      11      12
%   run count        4      14      24      84     165     272
%   longest run     13      73     159     687    1476    3084
%   run height    1158    4923    8805   30481   56649   96847
%   peak excur      20      56      93     339     709    1279
%   flat detect     65      81     225     459     932    1383
%   flat excur     345    1725    3429    8705   21392   33869

\begin{table}[ht]
\begin{center}
\caption{\label{tbl:Rperf} Timing of the R modality analysis.}
{\small
\begin{tabular}{lrrrrrr}
  & & & & & & \\
\multicolumn{1}{r}{ sample size } & 10k & 50k & 100k & 250k & 500k & 750k \\
  & & & & & & \\
Data Preparation \\
\qquad spacing             & $  0.00 \times $ & $  0.75 \times $ &     4 ms
                           & $  3.75 \times $ & $  6.00 \times $
                           & $  7.00 \times $ \\
\qquad interval spacing    & $  0.00 \times $ & $  0.50 \times $ &     4 ms
                           & $  4.25 \times $ & $  2.50 \times $
                           & $  5.75 \times $ \\
\qquad low-pass filtering  & $  0.01 \times $ & $  0.25 \times $ &   870 ms
                           & $  6.14 \times $ & $ 24.45 \times $
                           & $ 54.91 \times $ \\
\qquad mid-quantile        & $  0.00 \times $ & $  0.33 \times $ &     6 ms
                           & $  2.33 \times $ & $  6.00 \times $
                           & $  6.33 \times $ \\
Low-Pass Spacing Analysis \\
\qquad peak detection      & $  1.00 \times $ & $  0.50 \times $ &     2 ms
                           & $  2.00 \times $ & $  5.00 \times $
                           & $  6.00 \times $ \\
\qquad peak height test    & $  2.00 \times $ & $  1.00 \times $ &     0 ms
                           & $  1.00 \times $ & $  1.00 \times $ 
                           & $  1.00 \times $ \\
\qquad peak excursion test & $  0.11 \times $ & $  0.53 \times $ &   802 ms
                           & $  2.68 \times $ & $  6.01 \times $
                           & $  9.03 \times $ \\
\qquad flat detection      & $  0.19 \times $ & $  0.47 \times $ &   247 ms
                           & $  2.71 \times $ & $  5.89 \times $
                           & $ 10.34 \times $ \\
\qquad flat length test    & $  1.00 \times $ & $  1.00 \times $ &     0 ms
                           & $  1.00 \times $ & $  1.00 \times $ 
                           & $  1.00 \times $ \\
\qquad flat excursion test & $  0.10 \times $ & $  0.51 \times $ &  3209 ms
                           & $  2.60 \times $ & $  5.96 \times $
                           & $  9.84 \times $ \\
 Interval Spacing Analysis \\
\qquad peak detection      & $  0.13 \times $ & $  0.75 \times $ &     8 ms
                           & $  1.13 \times $ & $  1.38 \times $
                           & $  1.50 \times $ \\
\qquad run count test      & $  0.17 \times $ & $  0.58 \times $ &    24 ms
                           & $  3.50 \times $ & $  6.88 \times $
                           & $ 11.33 \times $ \\
\qquad longest run test    & $  0.08 \times $ & $  0.46 \times $ &   159 ms
                           & $  4.32 \times $ & $  9.28 \times $
                           & $ 19.40 \times $ \\
\qquad run height test     & $  0.13 \times $ & $  0.56 \times $ &  8805 ms
                           & $  3.46 \times $ & $  6.43 \times $
                           & $ 11.00 \times $ \\
\qquad peak excursion test & $  0.22 \times $ & $  0.60 \times $ &    93 ms
                           & $  3.65 \times $ & $  7.62 \times $
                           & $ 13.75 \times $ \\
\qquad flat detection      & $  0.29 \times $ & $  0.36 \times $ &   225 ms
                           & $  2.04 \times $ & $  4.14 \times $
                           & $  6.15 \times $ \\
\qquad flat excursion test & $  0.10 \times $ & $  0.50 \times $ &  3429 ms
                           & $  2.54 \times $ & $  6.24 \times $
                           & $  9.82 \times $ \\
\end{tabular}
}
\end{center}
\end{table}

\section{\rpkg{DimodalC} and {\rpkg DimodalPy} Ports}
\label{sec:ports}

\rpkg{DimodalC} is a C program for studying modality by using the
feature detectors and significance tests in the R reference version.
It forms the core of \rpkg{DimodalPy}, a Python wrapper that can be
used interactively.  Both are available in a source distribution
\citep{kreider26}.  The port simplifies several of the R functions,
for example limiting the Markov chain model for the longest run test
to 2 or 3 symbols.  It removes many of the utility functions,
automatically incorporating them into the analysis, and replaces
every R base function, notably for the permutation and excursion
tests.  It does not support changepoints, does not include the
\code{Ditrack()} function and class, and cannot set different
detector and test parameters for the low-pass and interval spacing.

\subsection{\rpkg{DimodalC} Interface}
\label{sec:Capi}

From the command line call
\begin{console}
C> DimodalC [-p <options file>]+ [<data file>] [-o <results>] [-g <plot>]
\end{console}
using text files for all arguments.  The data file, or stdin if not
provided, contains integers or floating point numbers separated by
whitespace or a comma or period, whichever is not the decimal separator.
Comments begin with a hash or semicolon, extend to the end of the line,
and are ignored.  Strings that cannot be parsed as numbers are skipped.
The options file contains key-value pairs separated by whitespace, a
colon, or an equals sign.  Key names follow the R version.  If the
command line contains more than one "-p" argument the files are read
in order, overwriting previous values.  This allows defining one set of
standard parameters and a second set for trying variants, and replaces
\code{Diopt.local()}.  If an options file does not exist then it is
created and filled with the current values.

The program prints to stdout the same feature and test result tables
as the R version, including matching low-pass to interval features
if both exist.  The matching uses two new options to define the
separation between peaks and overlap of flats.  The "-g" argument
causes \rpkg{DimodalC} to plot the spacing and features in SVG format.
The "-o" argument dumps all data, spacing, features, and test results
to a file in JSON format.  This output is compatible with a protobuf
specification provided in the source distribution, which also contains
a C and a Java program that reads back the output.

\subsection{\rpkg{DimodalPy} Interface}
\label{sec:pyapi}

The top-level Python function
\begin{console}
>>> import dimodalPy.dimodalPy as dm
>>> m = dm.check_modality(x, o)
\end{console}
returns a read-only instance $ m $ of the \code{"ModalAnalysis"} class.
Unlike the R package which provides a global persistent options
database, the Python implementation uses a local instance $ o $ of the
\code{"DiOpt"} class based on the underlying C data structure.  This
instance must be created and set to the default values before reading
a parameter file or setting internal options.  
\begin{console}
>>> o = dm.DiOpt()
>>> o.defaults()
>>> o.read(<options file>)
>>> o.set(key1=val1, key2=val2, ...)
\end{console}
The \code{set()} method can also take a dictionary of key/value pairs
as an argument.  The options file is the same as the C argument and
overrides existing values.  The class also has a \code{write(<file>)}
method to save the options.

As in R $ m $ contains members that depend on the analysis and tests
that have been run.  \code{m.data} is an instance of the \code{"DiData"}
class that holds the raw data, the spacing and its filtered versions,
and intermediate values for the runs and excursion tests.  The features
are captured in \code{m.lppeak} and \code{m.diwpeak}, instances of the
\code{"DiPeak"} class, and \code{m.lpflat} and \code{m.diwflat}, of
\code{"DiFlat"}.  These three classes are really containers of a lower
set of data structures that store the location and test statistics and
results for each feature, allocated as needed.  We use a consistent
index pair to retrieve individual elements, where the first element
specifies what to access --- \code{"row"} or \code{"stats"} for the
data, \code{"peak"} or \code{"extremum"} (also \code{"mm"} for
minimum-maximum) or \code{"peakmm"}, and \code{"flat"} for flats ---
and the second a numeric index into the array.  For the data there
are constants defined for this second element for each type.  For
example, \code{Row\_Di} addresses the spacing, \code{Row\_lp} the
low-pass spacing, and \code{Row\_Diw} the interval spacing.  As
examples,
\begin{console}
>>> row = m.data["row", dm.Row_Di]    # spacing
>>> m.data["stats", dm.Row_Di]        # mean, stdev, and range of spacing
>>> m.lppeak["mm", 1]                 # second extremum (location)
>>> m.diwpeak["peak", 0]              # first peak (test statistics/results)
>>> m.diwpeak["peakmm", 0]            # extremum for first peak (location)
>>> m.lpflat["flat", 2]               # third flat (location and tests)
\end{console}
Unlike R the port separates the extrema from the subset of peaks.
The first stores locations and raw values, the second test statistics
and results.  The data row is not itself a list of values.  Use a
separate index to access a single element, or \code{retrieve()} to
generate the list.  The row has the length of the original data.  It
sets invalid entries to NaN.  \code{stID} and \code{endID}, inclusive,
bound the valid indices for the row.
\begin{console}
>>> row[100]                         # spacing at index 100
>>> di = row.retrieve()              # entire row, with NaN at index 0
>>> di[di.stID:di.endID]             # only valid entries along row
\end{console}

\code{"ModalAnalysis"} has a \code{print()} method to show all the
results, the same as the C output.  \code{plot()} draws the graphs
using the \rpkg{matplotlib} library.

\subsection{Implementation}
\label{sec:CPyimpl}

The feature detectors follow the reference implementation after
defining new data structures to hold the results, replacing the R
SEXP containers.  Some small changes follow.  The data is always
on the original grid and not the valid subset, so locations do not
need to be shifted afterward.  The runs detector builds a run
length encoding and takes as a run value the average over the
series rather than the first point.  This may change the height of
a peak.  The peak detector omits the first and last extrema at the
ends of the data, which simplifies the edge cases in the R version
but may change the merging of peaks at the ends.  It finds the
support by scanning the data and including those points within the
height drop, while the R version worked with the runs representation.
The support range may shift by a few data points.  The flats
detector is unchanged.

Tests also have small changes compared to the reference version.
The height and length models do not change.  The runs count and
longest run tests are no longer generic, handling only the two or
three symbols found in the signed difference, and they do not
store additional information like the expected count distribution
parameters or the Markov transition and stationary matrices.  The
permutation test implements a shuffling function since the R
library is not available, and uses a different algorithm to
alternate three symbols, which had been split between C and R.
The run height probability will not match the reference results
exactly.

The Python version uses SWIG \citep{swig} to generate class
wrappers to the C data structures, as described in 
Section~\ref{sec:pyapi}.  All of the data structures except
\code{DiOpt} hold memory allocated during the analysis.  They are
therefore made read-only, with memory management done in C when
the analysis runs; we replace the contents of the SWIG destructor
for the \code{ModalAnalysis} instance returned by a call to C.
We also make \code{DiOpt} read-only and provide the \code{read()}
and \code{set()} methods to change members.  The latter uses the
same validation code as the former to ensure options have correct
values, and allows replacing bit flags with string names.  The
\code{plot()} method for the results is written in Python.
Additional C functions are needed for the indexing
\code{\_\_getitem\_\_()} and string conversion
\code{\_\_repr\_\_()} and \code{\_\_str\_\_()} methods.
\rpkg{DimodalPy} uses \rpkg{numpy} and SWIG to convert the data
into a flat memory structure for C.

\subsection{Performance}
\label{sec:Cperf}

Table~\ref{tbl:Cperf} shows the timing of the \rpkg{DimodalC}
program.  The analysis set-up and data are the same used in
Section~\ref{sec:Rperf}, although the subsets for each of the
25 repetitions will differ because the sampling function in
the R base library is not available.  The low-pass filtering
in the data preparation phase still scales as $ O( n^2 ) $,
although the base time is smaller.  Flat detection in the
low-pass spacing and the longest run test have worse than
linear scaling, but the other analysis steps are roughly
linear, except the models and peak detection in the interval
spacing which run in constant time.  The base time of the run
height permutation test has dropped by a factor of five, and
the peak excursion tests by a factor of two.  The performance
of the flat detector and excursion test do not change.

% Using time_Dimodal with kirkwood_all and 25 reps.
%  total feature counts and average flat length over all reps
%    LP:  peak  flat      len     Diw:  peak  flat      len
%  50k      48    25    46510             40    25    49223
% 100k      50    25    92055             31    25    98176
% 250k      50    25   232215             48    25   246368
% 500k      50    25   467088             40    25   492968
% 750k      50    25   702468             50    25   739475

\begin{table}[ht]
\begin{center}
\caption{\label{tbl:Cperf} Timing of the C modality analysis.}
{\small
\begin{tabular}{lrrrrr}
  & & & & & \\
\multicolumn{1}{r}{ sample size } & 50k & 100k & 250k & 500k & 750k \\
  & & & & & \\
Data Preparation \\
\qquad total               & $  0.26 \times $ &   235 ms     & $  6.05 \times $
                           & $ 23.94 \times $ & $ 54.14 \times $ \\
Low-Pass Spacing Analysis \\
\qquad peak detection      & $  0.66 \times $ &     1 ms     & $  2.23 \times $
                           & $  4.10 \times $ & $  8.83 \times $ \\
\qquad peak height test    & $  1.07 \times $ &     0 ms     & $  1.03 \times $
                           & $  1.11 \times $ & $  1.14 \times $ \\
\qquad peak excursion test & $  0.52 \times $ &   413 ms     & $  2.47 \times $
                           & $  5.53 \times $ & $  9.00 \times $ \\
\qquad flat detection      & $  0.30 \times $ &   542 ms     & $  3.87 \times $
                           & $ 11.12 \times $ & $ 27.59 \times $ \\
\qquad flat length test    & $  1.22 \times $ &     0 ms     & $  1.08 \times $
                           & $  0.99 \times $ & $  0.94 \times $ \\
\qquad flat excursion test & $  0.51 \times $ &  3152 ms     & $  2.50 \times $
                           & $  5.88 \times $ & $  9.38 \times $ \\
 Interval Spacing Analysis \\
\qquad peak detection      & $  0.67 \times $ &     8 ms     & $  0.93 \times $
                           & $  0.69 \times $ & $  0.66 \times $ \\
\qquad run count test      & $  0.48 \times $ &     0 ms     & $  2.53 \times $
                           & $  5.16 \times $ & $  7.96 \times $ \\
\qquad longest run test    & $  0.33 \times $ &   164 ms     & $  3.16 \times $
                           & $  9.95 \times $ & $ 18.66 \times $ \\
\qquad run height test     & $  0.42 \times $ &  1313 ms     & $  2.67 \times $
                           & $  5.70 \times $ & $  8.65 \times $ \\
\qquad peak excursion test & $  0.63 \times $ &    66 ms     & $  2.68 \times $
                           & $  5.84 \times $ & $  9.12 \times $ \\
\qquad flat detection      & $  0.31 \times $ &   191 ms     & $  2.46 \times $
                           & $  6.18 \times $ & $  6.04 \times $ \\
\qquad flat excursion test & $  0.51 \times $ &  3361 ms     & $  2.49 \times $
                           & $  6.38 \times $ & $  9.56 \times $ \\
\end{tabular}
}
\end{center}
\end{table}

\section{Kirkwood gaps example}
\label{sec:ex}

The package includes the dataset \rpkg{kirkwood}, a small subset of
the asteroid ephemeris maintained by Lowell Observatory
\citep{moskovitz22}.  The data is a vector of the semi-major axes of
asteroids within the orbit of Jupiter and with a known diameter.  In
total there are 2093 bodies.  These distances from the sun, in
astronomical units (AU), are interesting because both the modes and
anti-modes within have physical meaning.  The modes help identify
families of asteroids, bodies that formed together, perhaps as
remnants of a larger asteroid, and have similar orbits.  Their
composition and other orbital parameters, including eccentricity
and inclination, are also needed to tie the group to a common
ancestor.  The anti-modes correspond to the Kirkwood gaps, where
regular and repeated perturbation by Jupiter moves asteroids into
different orbits.  This occurs when its orbital period is a simple
fraction of that of Jupiter \citep{tsiganis02}.  The full ephemeris
has many asteroids between 4.266 and 5.0~AU, the latter the cut-off
for Jupiter's orbit, but they do not have measured diameters and
are not included in the dataset.

This data set is large and even the interval spacing is smooth because
the spacing is locally dense and consistent.  The values are taken to
six decimal places so ties are not a problem.  The features are small,
but the smooth filtered signal supports tightening the detector
parameters, allowing smaller peak heights and ripples for flats.  By
trial and error we adopt as default parameters
\begin{console}
R> opt <- Diopt(peak.fht=0.015, flat.fripple=0.0075)
\end{console}
The spacing at the start and end of the data is a factor 10 larger
than in the middle, and these relative detector parameters are
sensitive to the scale.  It may have been better to define heights
in terms of more robust statistics than the range of the spacing.

\begin{figure}
\centering
\begin{minipage}[t]{\textwidth}
\includegraphics[width=\textwidth]{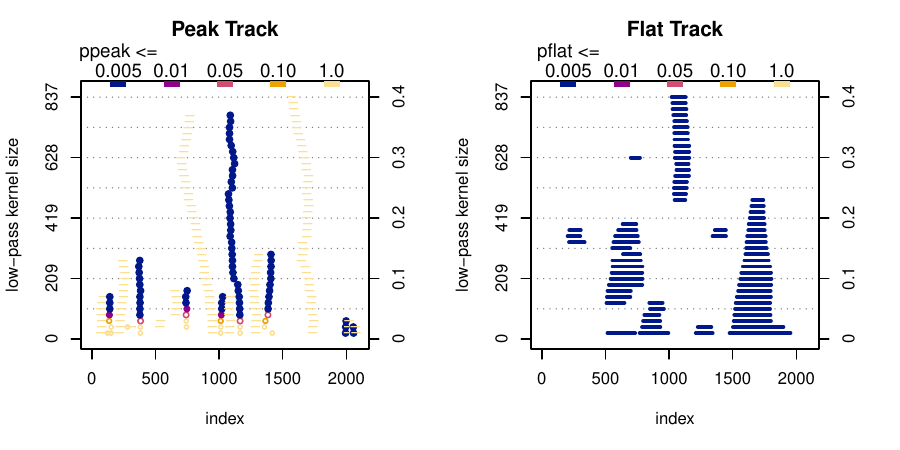}
\caption{\label{fig:trk} Position and probability of low-pass spacing
 features as kernel size changes.  Dashes in the left graph mark minima. }
\end{minipage}
\end{figure}

A tracking plot (Figure~\ref{fig:trk}) shows the position of peaks and
flats as the filter or interval size changes.
\begin{console}
R> trk.lp <- Ditrack(kirkwood, "lp")
R> plot(trk.lp)
\end{console}
The color of the points and bars indicate the best probability of the
feature, and dots are filled in if any test passes its acceptance
level.  Such peaks appear at the 5\% kernel size, along with two
flats.  The peaks fade away as the filter widens, leaving one in the
middle when $ f_{lp} = 15\% $.  Flats exist to each side of this
central peak; the left shifts 500 points to the left at
$ f_{lp} = 7\% $ when the peak near index 750 disappears.  The
central peak widens as the filter grows, eventually supporting a
flat at $ f_{lp} = 23\% $.  If we had used the default feature
parameters, the mode tree would not have included the third and
fourth peaks and the flats would be much larger.  From a similar
plot for the interval spacing the best interval width is also 5\%.

For reproducible results from the analysis we will not use the
changepoint detectors and will set RNG seeds for the excursion and
runs permutation tests.  The options here add to those already set.
Marking flats with bars instead of boxes is easier to see with
dense data.
\begin{console}
R> opt <- Diopt(analysis=c("lp", "diw"), lp.window=0.05, diw.window=0.05,
+               excur.seed=3, perm.seed=5, mark.flat="bar")
R> m <- Dimodal(kirkwood)
R> plot(m)
\end{console}

\begin{figure}
\centering
\begin{minipage}[t]{\textwidth}
\includegraphics[width=\textwidth]{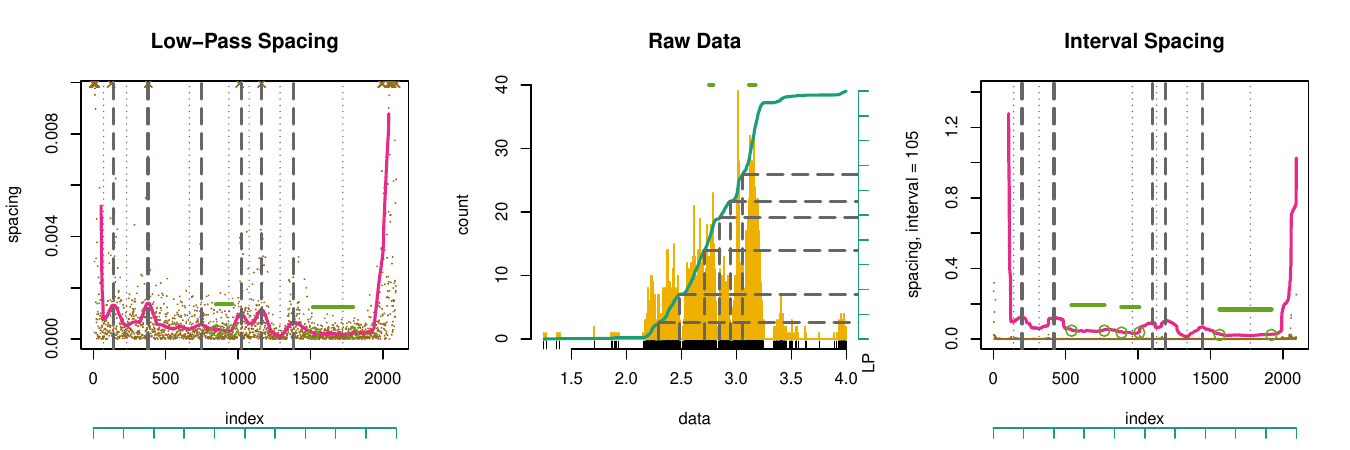}
\caption{\label{fig:kg} Low-pass and interval spacing analysis of the
kirkwood data.}
\end{minipage}
\end{figure}

Figure~\ref{fig:kg} shows the results.  There are three graphs, with
the low-pass spacing analysis to the left and interval spacing to
the right, and a histogram and rug plot in the middle.  The curve
atop the histogram is the distribution.  The axis to the right and
those below the spacing graphs mark the deciles and help convert
indices in the spacing to data values.  Peaks are marked with dashed
lines and minima with dotted.  Flats are marked with horizontal lines
above the data, or at the top of the histogram.  Significant features
use heavy lines.  The features marked in the histogram come from the
low-pass spacing.  Changepoints, were they generated, would be marked
with long ticks at the top of the graphs.

The basic \code{data} member of the result is a matrix with at least
four rows, \code{"x"} the source, \code{"xsort"} the order statistics
or sorted data, \code{"Di"} the spacing with NA at index 1, and
\code{"xmid"} the mid-quantile approximation.  The mid-distribution
just reproduces the original values because the Kirkwood data has no
ties.  The low-pass analysis adds a row \code{"lp"}, and the interval
spacing rows \code{"Diw"} and \code{"signed"}, the latter used for
the runs testing.  Invalid data, those points not fully covered by
the low-pass kernel or interval spacing, or lost by the differencing
in the spacing and runs data, are set NA, and attributes attached to
the matrix give the valid indices.  Printing this member of $ m $
gives a summary of the data.

\begin{console}
R> m$data

  data source        kirkwood 
  low-pass spacing   with 105 (0.050) kaiser filter
  interval spacing   with 105 (0.050) interval
     positions at interval end; shift by -52 vs. low-pass

Information
  row         valid       range      sd     
  x           1 - 2093    3.293      0.361  
  Di          2 - 2093    0.3187     0.01331
  LP Di      54 - 2041    0.09225    0.00235
  Diw       106 - 2093    1.263      0.1143
\end{console}

The low-pass spacing has six peaks and two flats.  The interval
spacing has five peaks and three flats.  We use the utility function
\code{match\_features()} to identify those that appear in both.  Peaks
match if they are close, by default within 10 points but here 30
because the data is large.  Flats match  if their common segment is
some fraction of their lengths, by default 70\%.
\begin{console}
R> match.features(m, near=30)

  matching peaks within  30  points after centering intervals
           low-pass                      interval       
     pos      raw    #pass         pos      raw    #pass
     141   (2.296)       1         198   (2.304)       1
     378   (2.487)       1         419   (2.471)       1
    1021   (2.848)       1        1101   (2.864)       1
    1162   (2.944)       1        1189   (2.921)       2
    1385   (3.056)       1        1444   (3.061)       3

  matching flats each overlapping by 0.70 after centering intervals
                 low-pass                              interval               
        pos           raw       #pass         pos           raw       #pass
     850 -  959  (2.755 2.789)      1      886 - 1008  (2.749 2.789)      1
    1519 - 1793  (3.118 3.173)      1     1564 - 1921  (3.116 3.191)      1
\end{console}
The tables give the positions or indices of the feature, the
corresponding data value from \code{"xmid"}, and the number of tests
passing their acceptance levels.

The spacing finds gaps in the asteroid belt at 2.30, 2.48, 2.86,
2.93, and 3.06~AU.  These are averages of the matching peaks, which
we generate using the utility function \code{select.peaks()} to
separate the actual peaks from all extrema and by dropping the third
low-pass peak which does not have a match.
\begin{console}
R> a_gap <- (select.peaks(m$lp.peaks)$x[-3L] +
+            select.peaks(m$diw.peaks)$x) / 2
\end{console}

The resonance of a gap with Jupiter is the ratio of the corresponding
orbital periods.  Kepler's Third Law says $ p^2 \propto a^3 $ for 
period $ p $ and orbital axis $ a $; the proportionality constant will
drop out when we take the ratio.  The resonance is
\begin{equation} \label{eq:res}
\left( \frac{a_{Jupiter}}{a_{gap}} \right)^{3/2}
\end{equation}
where the axis of Jupiter is 5.201~AU.  The gaps have resonances of
3.40, 3.04, 2.46, 2.36, and 2.22.  These correspond to ratios of 7:2
or 10:3, 3:1, 5:2, 7:3, and 9:4.

Bounding each peak are the minima and a support range at some fraction
of the height, by default 90\%.  Similar to Full-Width-at-Half-Maximum,
the support better captures the peak's width when the minima lie
within a long flat, for example the last minimum at index 1776.  It
determines the draw size for the excursion test.  Runs tests extend
the peak to the minima.  For the interval spacing
\begin{console}
R> m$diw.peaks

  location of maxima 
  pos       raw          minima pos (raw)          support pos
   198   (2.304)     141 -  316 (2.237 - 2.397)     146 -  262
   419   (2.471)     316 -  960 (2.397 - 2.773)     337 -  873
  1101   (2.864)     960 - 1128 (2.773 - 2.884)    1003 - 1119
  1189   (2.921)    1128 - 1339 (2.884 - 3.016)    1145 - 1303
  1444   (3.061)    1339 - 1776 (3.016 - 3.160)    1352 - 1587

  statistics
  pos     hexcur   nrun     runlen   runht                     
   198    0.05697   89        9       36                        
   419    0.08053  294       12       87                        
  1101    0.05516   78       10       38                        
  1189    0.08004   94       16       77                        
  1444    0.04815  178       22       72                        

  probabilities
  pos     pexcur   pnrun    prunlen  prunht    ppeak    pass   
   198    0.3387   0.5693   0.1944  _0.0000_   0.0006   T (1)  
   419    0.8104   0.0206   0.1577  _0.0029_   0.0029   T (1)       
  1101    0.3547   0.1809   0.1392  _0.0002_   0.0002   T (1)  
  1189    0.3799   0.2621  _0.0079_ _0.0000_   0.0000   T (2)  
  1444    0.6656  _0.0000_ _0.0006_ _0.0000_   0.0000   T (3)  

  accept at
          0.0100   0.0100   0.0100   0.0050
\end{console}
Underscores surrounded accepted test results; on the screen these
become underlines.  The pass column is marked T(rue) if at least one
test meets its acceptance level.  The peak at 1444 passes all three
runs tests, while that at 1189 passes the longest run and run height
permutation tests.  The other peaks pass the permutation test.  No
peak passes the excursion test; the large spacing in the tails,
particularly the trailing, when added to the draw pool routinely
leads to large re-constructed features, making the actual peaks
appear unremarkable.

All of these peaks in the low-pass spacing pass the height model
test, but none the excursion test.  There is a sixth peak at 747
(2.710) that the height model accepts with probability 0.0010, but
again not the excursion test.  Its resonance is 2.66, a ratio of 8:3.

The low-pass flats pass the excursion test but not the length model.
The problem is that the model has been pushed outside its bounds by
the size of the data; the 0.05 critical length from the model is
more than the available data.  Using the liberal Weibull base
distribution would still give a critical length over half the data.
The first mode seems to correspond to the Ceres family of asteroids
at a distance of 2.76~AU, while the second includes the Themis family
at 3.13~AU \citep{kozai79} \citep{gradie79}.  These identifications
are tentative as they consider only the semi-major axis for grouping.
The interval spacing has a third significant flat between 2.59 and
2.69~AU, passing the excursion test with probability 0.000 (rounding
to three decimal places).  This range includes the Eunomia and
Prosperina families.

\begin{console}
R> m$lp.flats

                                statistics
  endpoints          raw        len      hexcur                       
   850 -  959  (2.755 2.789)    110      6.239e-05                    
  1519 - 1793  (3.118 3.173)    275      6.543e-05                    

                                probabilities
  endpoints                     plen     pexcur       pflat    pass   
   850 -  959                   0.9626  _0.0003_      0.0003   T (1)  
  1519 - 1793                   0.9064  _0.0000_      0.0000   T (1)  

  accept at                     0.0500   0.0100
\end{console}

The histogram and rug plot in Figure~\ref{fig:kg} show two clusters
of asteroids beyond 3.0~AU.  They generate peaks that are visible in
the lower right corner of Figure~\ref{fig:trk}.  If we decrease the
filter and interval size to 31 points, a fractional width of 0.015,
then peaks appear in both spacings, at 3.36 and 3.60~AU.  These are
resonances of 1.93 and 1.74, with ratios near 2:1 and of 7:4.  The
inner gaps are no longer accepted and the peak at 3.06~AU disappears.
New flats appear at 2.58, 2.64, and 3.01~AU as the larger spacing
increases the ripple range.  Using \code{Diopt.local} to temporarily
change the analysis parameters,
\begin{console}
R> m2 <- Dimodal(kirkwood, Diopt.local(lp.window=0.015, diw.window=0.015,
+                                      peak.fht=0.025, peak.relht=0.10))
R> match.features(m2, near=30)

  matching peaks within  30  points after centering intervals
           low-pass                      interval       
     pos      raw    #pass         pos      raw    #pass
     152   (2.319)       0         149   (2.292)       0
     382   (2.515)       0         391   (2.477)       1
    1012   (2.833)       0        1034   (2.846)       1
    1166   (2.965)       0        1183   (2.965)       1
    1995   (3.348)       1        2015   (3.368)       1
    2056   (3.633)       2        2068   (3.565)       1

  matching flats each overlapping by 0.70 after centering intervals
                 low-pass                                interval               
        pos           raw       #pass         pos           raw       #pass
     411 -  520  (2.548 2.609)      1      422 -  539  (2.544 2.610)      1
     543 -  654  (2.614 2.666)      1      549 -  673  (2.613 2.667)      1
     797 -  986  (2.735 2.799)      1      791 - 1003  (2.724 2.800)      1
    1216 - 1338  (3.004 3.022)      1     1227 - 1352  (3.002 3.022)      1
    1484 - 1907  (3.107 3.201)      1     1490 - 1960  (3.104 3.214)      1
\end{console}
    
The simulations in \citep{tsiganis02} explain the gaps at 3:1, 5:2, 2:1,
and 7:4.  7:3 appears to be a special case.  In these results the 2:1
resonance is mis-positioned, but this is because there are no asteroids
in the gap, which lies at 3.276~AU.  The spacing indeed shows increases
here, at data index 1993, with an obvious strong peak between the
neighbors; the spacings are 0.00044, 0.09225, and 0.00654.  However, a
second large spacing of 0.01311 at index 1998 pulls the low-pass peak
to index 1995, out of the actual gap.  The interval spacing peak also
lies to the side.  The outlier at 1998 combines in filtering with 1995
and shifts the peak by only a few data points, which distorts the
position because of the large spacing step.  With discrete data the
mid-quantile approximation would help correct the offset.

Were we to run changepoint detectors, we would find they lie at the
transition between mode and gap, but not directly at either feature.
The full suite of supported detectors agrees on changepoints to both
sides of three peaks, at indices 370 and 386, 1953 and 1997, and 2053
and 2087, averaging to distances of 2.50~AU, 3.28~AU, and 3.78~AU
respectively.  The changepoints have picked up the two outer clusters.
There are also changes to one side of a peak, at 1005 (2.811),
1209 (3.000), and 1345 (3.025).  A tenth changepoint at index 12 falls
in the initial tail.

Table~\ref{tbl:kirkwood} summarizes the peaks and gaps.  The pairs in
the table are the number of passing low-pass and interval tests.
Changepoints bounding a peak on both sides are marked with a star.

\begin{table}
\begin{center}
\caption{\label{tbl:kirkwood} Kirkwood gaps found.}
{\small
\begin{tabular}{clcrccrccrcc}
 & & & & & & & & & & & \\
 & & & & \multicolumn{5}{c}{filter size} & & & \\
 & & ideal & & \multicolumn{2}{c}{0.05} & & \multicolumn{2}{c}{0.015} &
 & \multicolumn{2}{c}{changepoint} \\
resonance & ratio & axis [AU] & & axis & \#pass & & axis & \#pass & &
  \multicolumn{2}{c}{axis [AU]} \\
3.40 & 7:2 or 10:3 & 2.25 or 2.33
                          & & 2.30 & (1, 1) & & 2.31 & (0, 0) & &      &   \\
3.04 & 3:1         & 2.50 & & 2.48 & (1, 0) & & 2.50 & (0, 1) & & 2.50 & * \\
2.46 & 5:2         & 2.82 & & 2.86 & (1, 1) & & 2.84 & (0, 0) & & 2.81 &   \\
2.36 & 7:3         & 2.96 & & 2.93 & (1, 2) & & 2.97 & (0, 0) & & 3.00 &   \\
2.22 & 9:4         & 3.03 & & 3.06 & (1, 3) & &      &        & & 3.03 &   \\
1.93 & 2:1 (gap)   & 3.28 & &      &        & & 3.36 & (1, 1) & & 3.28 & * \\
1.74 & 7:4         & 3.58 & &      &        & & 3.60 & (2, 1) & & 3.78 & * \\
 \\
 & & & & 
\multicolumn{8}{l}{\small \#pass are (low-pass, interval) spacing counts} \\
\end{tabular}
}
\end{center}
\end{table}

In the source distribution of \rpkg{DimdalCPy} you will find two
files with the setup and Kirkwood data.  From the command line run
\begin{console}
C> DimodalC -p kirkwood_param kirkwood_data -g kw.svg
\end{console}
This will print all of the feature tables, including their matching
between the low-pass and interval spacing, and will write a version
of Figure~\ref{fig:kg} to the SVG file.  Within Python,
\begin{console}
>>> import io ; import numpy as np ; import dimodalPy.dimodalPy as dm
>>> o = dm.DiOpt()
>>> o.defaults()
>>> o.read("kirkwood_param")
>>> f = open("kirkwood_data") ; s = f.read().replace(",", "") ; f.close()
>>> x = np.genfromtxt(io.StringIO(s.rstrip() + " nan")).flatten()
>>> x = x[0:(len(x)-1)]
>>> m = dm.check_modality(x, o)
>>> m.print()
>>> m.plot()
\end{console}
The processing of the data file requires removing the comma separators,
which would create an extra column on each line, and padding the last line
with an NaN to complete the table.  This dummy value is then removed.

\section*{Disclosure of Interest and Funding Statements}
There are no competing interests to declare.  No funding was received
for this work.

\bibliographystyle{tfcad}
\bibliography{Dimodal}

\end{document}